\PassOptionsToClass{12pt,pra,aps}{revtex4-1}
\documentclass[apjl]{emulateapj}
\usepackage{amsmath}
\usepackage{natbib}

\begin{document}
\bibliographystyle{plainnat} 

\title{Study of GRB light curve decay indices in the afterglow phase}

\author{Roberta Del Vecchio$^1$}
\email{E-mails: roberta@oa.uj.edu.pl (RDV)}

\author{Maria Giovanna Dainotti$^{2,1,3}$}
\email{mdainott@stanford.edu (MGD)}

\author{Micha\l{} Ostrowski$^1$}

\affil{$^1$ Astronomical Observatory, Jagiellonian University, ul. Orla 171, 30-244 Krak{\'o}w, Poland \\
$^2$ Physics Department, Stanford University, Via Pueblo Mall 382, Stanford, CA, USA\\
$^3$ INAF-Istituto di Astrofisica Spaziale e Fisica cosmica, Via Gobetti 101, 40129, Bologna, Italy}
\email{michal.ostrowski@uj.edu.pl (MO)}

\begin{abstract}
\noindent In this work we study the distribution of temporal power-law decay indices, $\alpha$, in the Gamma Ray Burst (GRB) afterglow 
phase,  
fitted for $176$ GRBs (139 long GRBs, 12 short GRBs {\it with extended emission} and 25 X-Ray Flashes (XRFs)) with known redshifts. 
These indices are compared with the temporal decay index, $\alpha_W$, derived with the light curve fitting using the 
\cite{willingale07} model. 
This model fitting yields similar distributions of $\alpha_W$ to the fitted $\alpha$, but for individual bursts a difference can 
be significant. Analysis of ($\alpha$, $L_a$) distribution, where $L_a$ is the characteristic luminosity at the end of the
plateau, reveals only a weak correlation of these quantities. However, we 
discovered a significant regular trend when studying GRB $\alpha$ values along the \cite{dainotti2008} 
correlation between $L_a$ and the end time of the plateau emission in the rest frame, $T_a^*$, hereafter LT correlation. 
We note a systematic variation of the $\alpha$ parameter distribution with luminosity for any selected $T_a^*$. 
We analyze this systematics with respect to the fitted LT correlation line, expecting that  
 the presented trend may allow to constrain the GRB physical models.
We also attempted to use the derived correlation of $\alpha(T_a)$ versus $L_a(T_a)$ to diminish the luminosity scatter
related to the variations of $\alpha$ along the LT distribution, a step forward in the effort of
standardizing GRBs. A proposed toy model accounting 
for this systematics applied to the analyzed GRB distribution results in a slight increase of the LT correlation coefficient. 
\end{abstract}

\section{Introduction}

Gamma Ray Bursts (GRBs) with their powerful  emission processes are observed up to high redshifts, $z > 9$ \citep{cucchiara11}. 
A significant progress in studying GRB observations has been the advent of \textit{Swift} satellite \citep{gehrels04}, 
which has 
revealed a more complex behavior of the light curves \citep{OB06,Nousek2006,Zhang2006,Sak07} than in the past.
There are several emission models proposed in the literature providing predictions for characteristic GRB light curve features. 
Well known is the model of \cite{meszaros98,meszaros99,meszaros06}, consisting of jet internal shocks generating the GRB prompt 
phase emission and external shocks of the GRB expanding fireball generating the afterglow emission.

In the present paper we study distributions of the GRB afterglow parameters versus light curve temporal decay indices $\alpha$, for the power-law decay observed in  
the afterglow phase, with the X-ray luminosity $L_a$. We analyze an extended sample of 176 GRBs with known 
redshifts observed by \textit{Swift} from January 2005 to July 2014. In the presented analysis we use the 
LT correlation \citep{dainotti2008}, updated in 
\cite{dainotti2010,dainotti11a,dainotti13,dainotti15a}, between the derived characteristic afterglow plateau luminosity, 
$L_a$, and time, $T_a^*$ (an index~* indicates the GRB 
rest frame quantity). An attempt to study similar afterglow properties was presented by \cite{Gendre2008}. 
They analyzed the ``late" light curve properties at the time of 1 day after the burst to reveal existence of grouping the
GRB luminosities into two groups which also differ in their redshift distributions. In their study they noted relations 
among some GRB parameters, in particular a non-trivial distribution of X-ray spectral indices versus the light curve 
temporal decay indices, but no dependence of these indices on the GRB luminosity. 
In addition, prompt-afterglow correlations have been studied by \cite{dainotti11b,dainotti15b}, \cite{margutti13} and \cite{grupe13}.

Importance of the present study results also from the fact that the afterglow LT correlation has already been the object of 
theoretical modeling either via accretion \citep{kumar08,lindner10,cannizzo09,cannizzo11}, via a magnetar model \citep{dallosso11, bernardini11, rowlinson12, rowlinson13, rowlinson14,rea15} or energy injection \citep{sultana13, leventis14, vaneerten14a, vaneerten14b}
and there were attempts to apply it as a cosmological tool \citep{cardone09,cardone10,dainotti13b,postnikov14}.
Here, we extend the LT correlation study looking into its possible dependence on the additional physical parameter $\alpha$ 
characterizing the afterglow light curve.

Below, in Section 2 we introduce the data set analyzed in this study. In Section 3 we describe the performed 
observational data analysis and the derived distributions of  decay indices $\alpha$.
The analysis reveals a weak correlation for $\alpha$ and  $L_a$, but a significant systematic trend along the correlated 
($L_a$, $T_a^*$) distribution. In Section 4 we present final discussion and conclusions. 
We shortly consider physical interpretation of the $\alpha$ distribution. Then, 
we preliminary explore a new possibility of using GRBs as cosmological standard candles, 
illustrated with a proposed toy model involving scaling GRB afterglow luminosity to the 
selected standard $\alpha_0$. 

The fitted slopes and normalization parameters of the correlations presented in this paper are derived using
the \cite{dagostini05} method. The $\Lambda$CDM cosmology applied here uses the parameters 
$H_0 = 71 \ km\ s^{-1} \ Mpc^{-1}$, $\Omega_{\Lambda}= 0.73$ and $\Omega_M= 0.27$~. 

\section{Data Sample}

Below, we analyze distribution of afterglow light curve decay indices, $\alpha$, for the data set of 176 GRBs
with known redshifts,  observed by \textit{Swift} from January 2005 to July 2014. Within this sample we consider separately 
subsamples of 139 long GRBs, 25 X-Ray Flashes (XRFs)\footnote{XRFs are bursts of high energy emission similar to long GRBs, but 
with a spectral peak energy one order of magnitude smaller than in the long GRBs and with fluence
greater in the X-rays than in the gamma ray band. Sometimes XRFs are considered to be misaligned long GRBs \citep{ioka2001,yamazaki02} and it is why 
we also analyze both these samples together.}, 12 short GRBs with extended emission. The sample of 164 long GRBs and XRFs is also considered together as a single sample.

As described in \cite{dainotti13}, all the analyzed light curves were fitted using an analytic functional form proposed by \cite{willingale07}. 
The considered sample was chosen from all \textit{Swift} GRBs with known redshifts by selecting only those events which allowed a reliable
afterglow fitting. 
The fits provided physical parameters for the GRB afterglows, 
including its characteristic luminosity and time, $L_a$ and $T_a^*$, at the end of the afterglow plateau phase, 
and the power-law temporal decay index, $\alpha_W$, for the afterglow decaying phase\footnote{These data are available upon
request from M.G. Dainotti.}. The fitted indices $\alpha_W$ are influenced by the requirement of the best 
global light curve fitting for the considered analytic model. 
Therefore, we decided to apply a different procedure for the derivation of the temporal decay index $\alpha$ 
to be used in the following analysis, intended to provide 
a more accurate fit of the light curve power-law decay part immediately after the plateau. In each GRB we selected the afterglow light
curve section with a power-law and we performed the $\chi^2$ fitting of $\alpha$ 
in such a range, as presented in the ON LINE MATERIAL including a set of figures for all GRBs 
showing the performed fits and providing the fitting parameters in the table. 
We compared these parameters with those by \cite{evans09} and the ones quoted in the Swift Burst Analyser 
(http://www.swift.ac.uk/burst$_{-}$analyser/), 
having nearly the same 
$\alpha$ values in the majority of cases. However, also significant differences are found in individual cases due 
to the different time ranges considered for the fits.

The applied procedure allowed to remove all clear deviations of the power-law from the fitting due to flares and 
non-uniformities in the observational data. In the case of a break in the decaying part of
the light curve, a value of $\alpha$ was fitted to the brighter/earlier part of the light curve. We have to note that in some rare 
cases it was impossible to decide if the first part of the light curve can be considered the decaying part, or still a steep 
plateau phase, and the presented fits can be disputed. 
We decided to use all derived data in the analysis and correlations studied in this paper, leaving possibility of some particular 
data selection and/or rejecting some events from the analysis to the future study.

\begin{figure}[htbp]
\centering
\includegraphics[width=0.976\hsize,angle=0,clip]{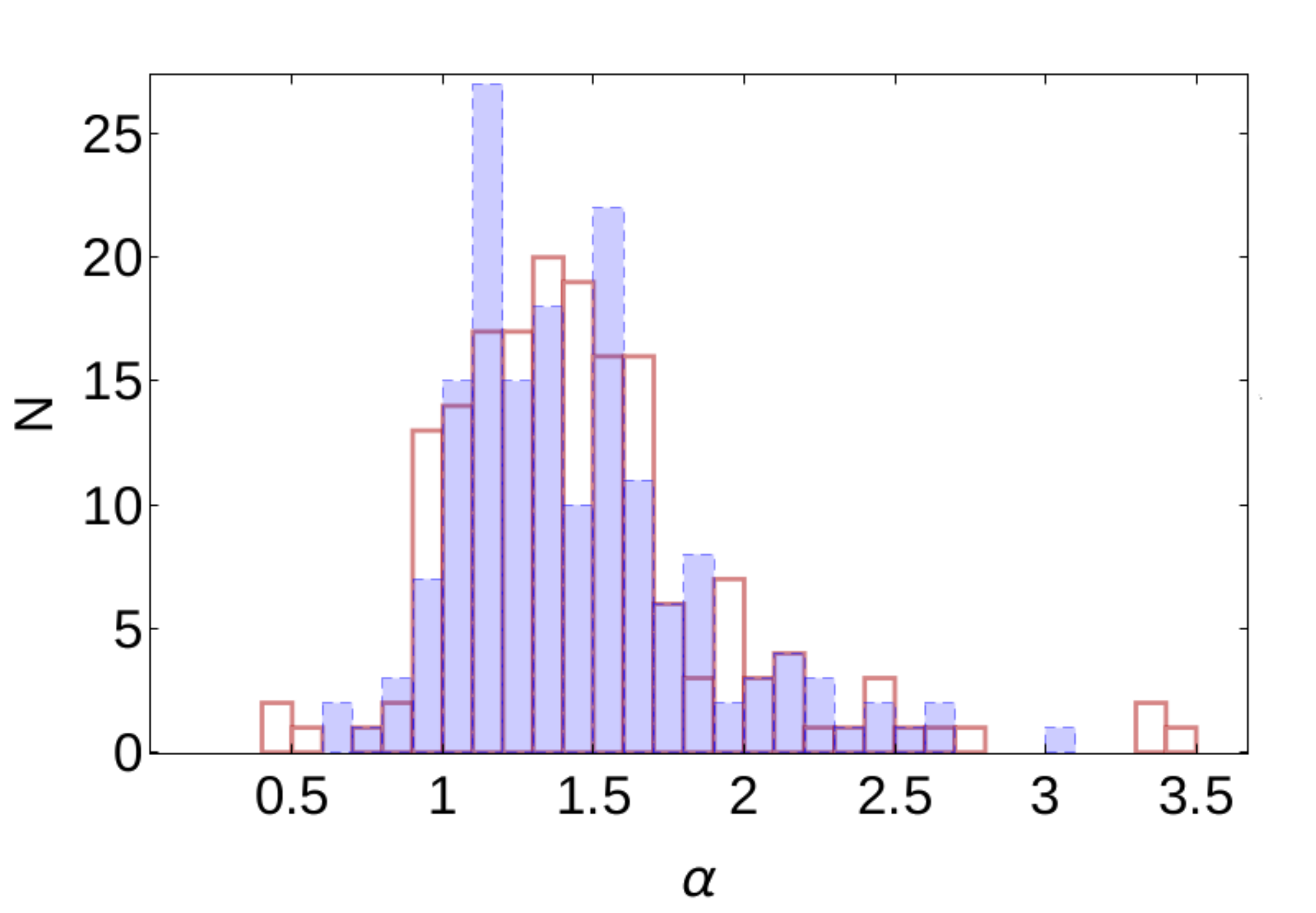}
\includegraphics[width=0.915\hsize,angle=0,clip]{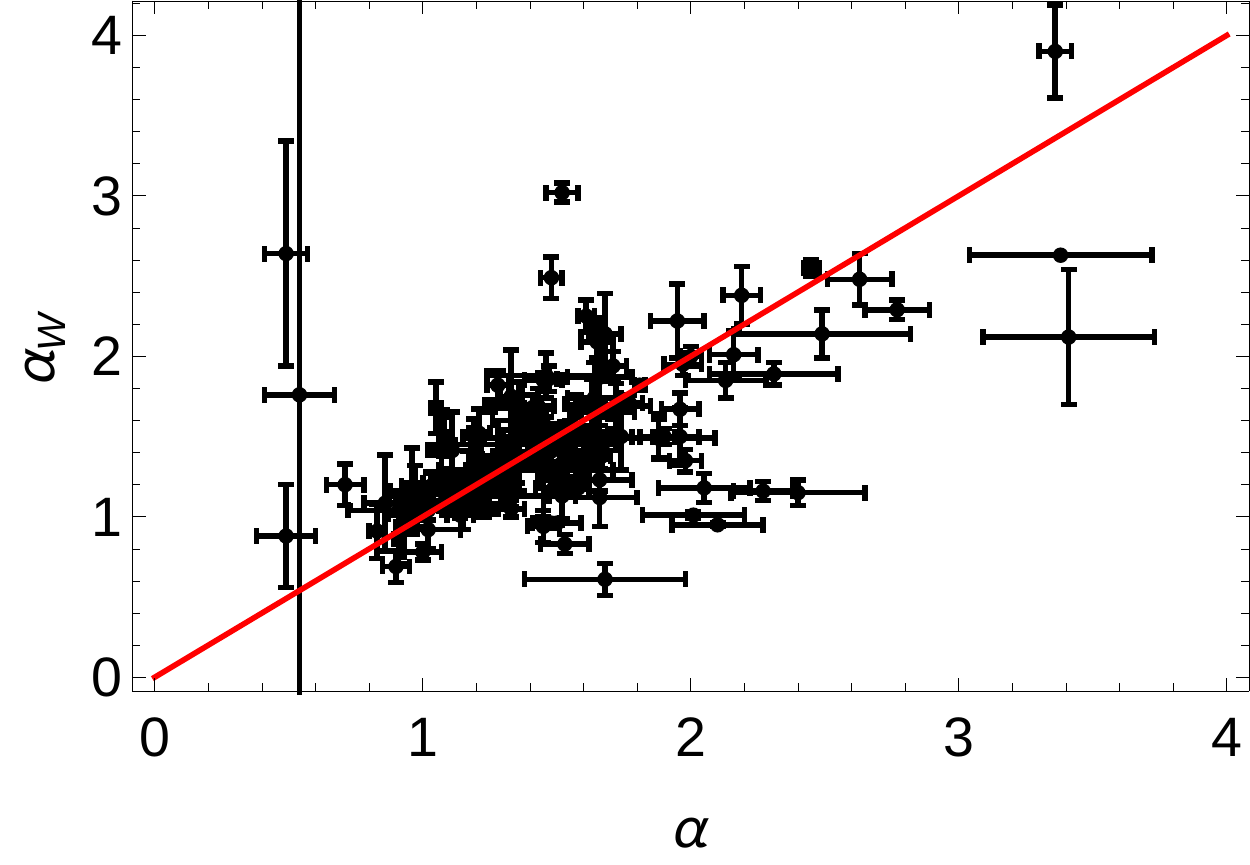}
\caption{\footnotesize {Comparision of the derived indices $\alpha$ and $\alpha_W$. 
Upper panel: histograms of fitted $\alpha$ (solid red line) and $\alpha_W$ (shaded area). 
Bottom panel: the $\alpha_W$ versus $\alpha$ distribution. The red line $\alpha=\alpha_W$ is provided for reference.}}
   \label{fig:1}
\end{figure}

Comparison of the $\alpha$ and $\alpha_W$ distributions in Fig. \ref{fig:1} shows that both measured 
decay indices have similar distributions, 
but differences for individual fitted values can be significant. The parameters of the Gaussian fits 
for both presented distributions are: a mean value $\mu \left( \alpha \right)=1.40$ with standard deviation 
$\sigma\left( \alpha \right)=0.46$ for our power-law fitting compared to $\mu\left(\alpha_W\right)=1.45$ and 
$\sigma\left(\alpha_W\right)=0.45$ for the Willingale model fitting.
The P-value of the T-test between these two distributions is 0.89, indicating no statistically significant 
differences between the two distributions.

\section{Analysis of the afterglow decay light curves}

\begin{figure}[htbp]
\centering
\includegraphics[width=1\hsize,angle=0,clip]{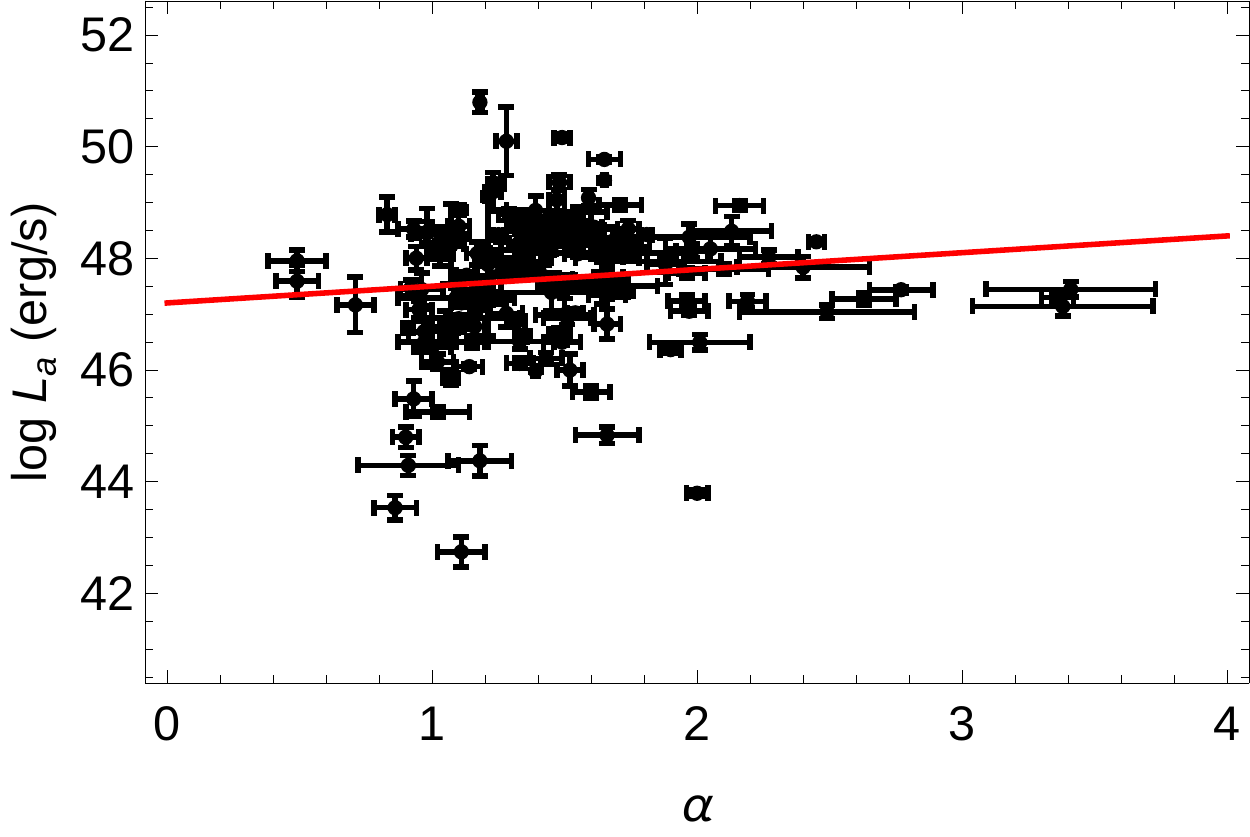}
\caption{\footnotesize Distribution of $\log L_a$ versus $\alpha$ for the long GRBs+XRFs subsample. 
The line presents a fitted weak correlation. }
   \label{fig:2}
\end{figure}

By adopting the analyzed subsample of long GRBs+XRFs, systematic trends in the ($\log L_a$, $\alpha$) distribution can be
studied.
As presented in Fig. \ref{fig:2}, there is an indication that these quantities are (weakly) correlated. 
However, scatter of points around the correlation line is substantial and the derived
Spearman \citep{spearman} correlation coefficient $\rho = 0.17$ is small. 
The fitted correlation line is $\log L_a$ = $0.30\,  \alpha$ + $47.20$, showing that {\em on average} faster light curve decay 
seems to occur for GRBs with higher luminosities. However, here we stress again, the trend is weak in this highly scattered 
distribution. The same analysis using the long GRBs subsample only leads to a slightly weaker correlation with
 $\rho=0.14$ and $P=10^{-4}$.

It should be noted that the large scatter of the GRB luminosity distribution can not be due only to the 
fitting method used for its derivation. Significant contribution to this scatter must 
result from the very nature of the GRB sources, possibly modified by the explosion geometry. 
 
\begin{figure}[htbp]
\centering
\includegraphics[width=1.0\hsize,angle=0,clip]{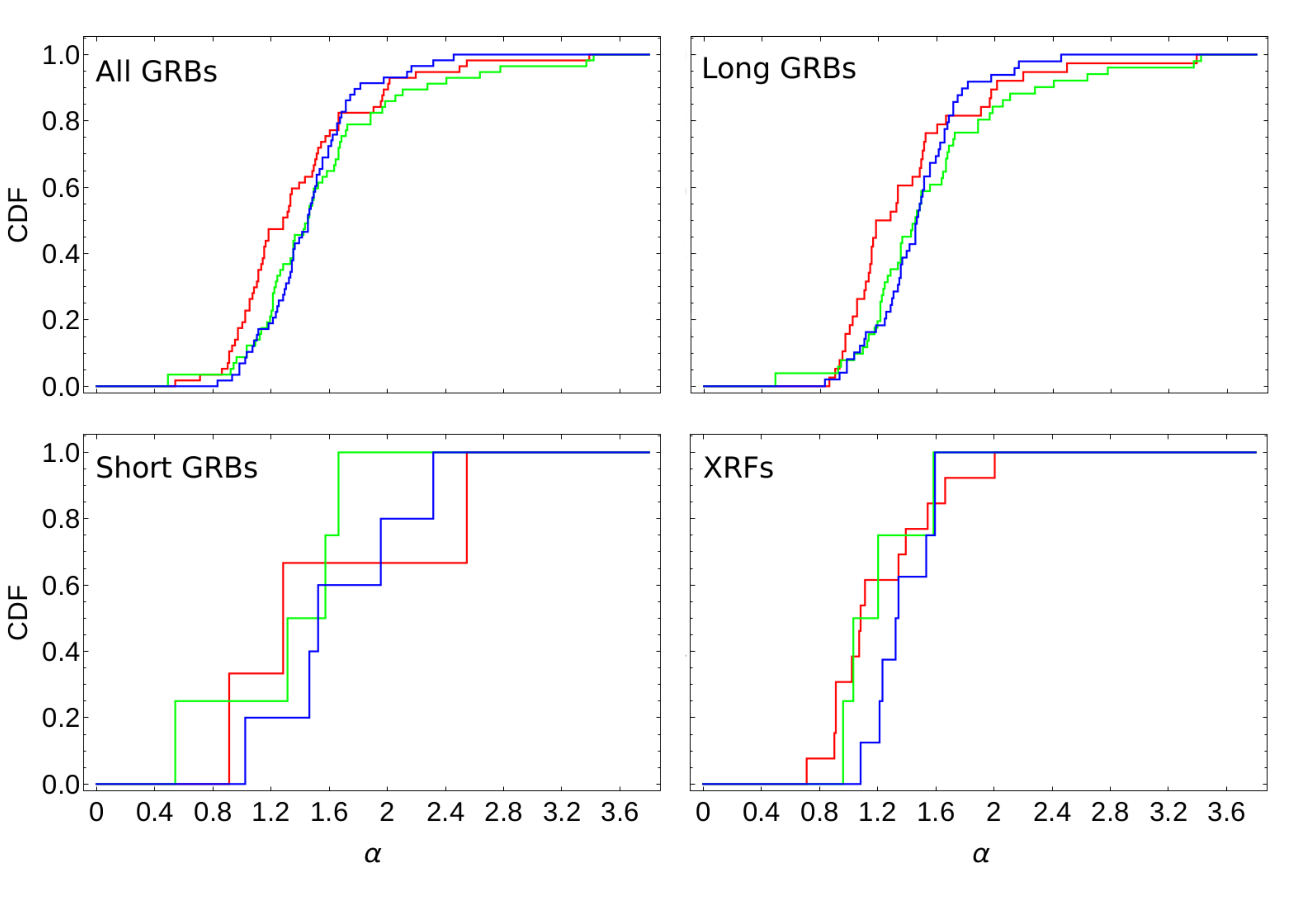}
\caption{\footnotesize Normalized cumulative distribution functions, $CDF$, versus $\alpha$ for the analyzed GRB subsamples 
in 3 considered luminosity ranges: $\log L_a < 47.25$ (red), $47.25 < \log L_a < 48.2$ (green) and $\log L_a > 48.2$ (blue).}
   \label{fig:3}
\end{figure}

\begin{figure}[htbp]
\centering
\includegraphics[width=1\hsize,angle=0,clip]{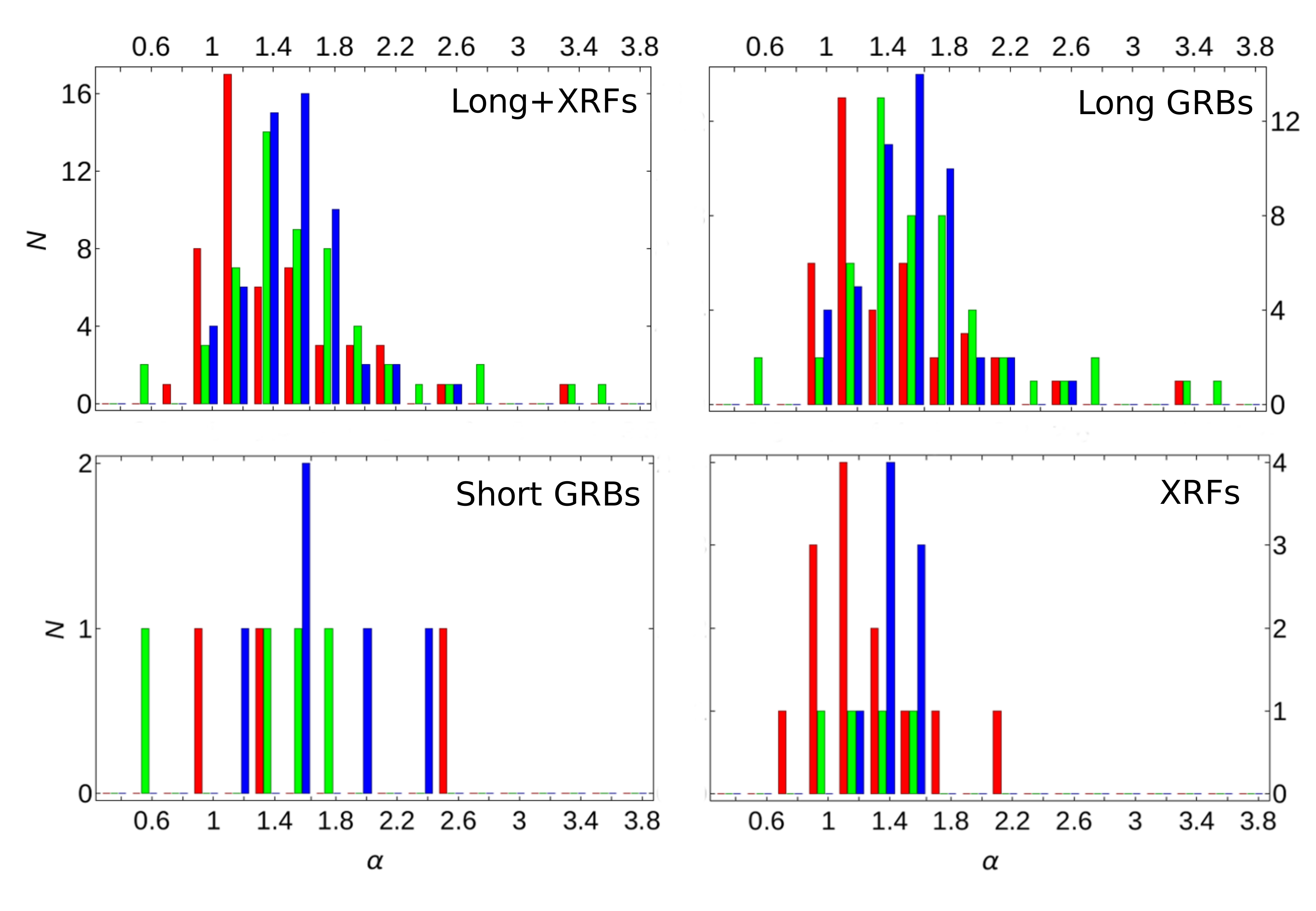}
\caption{\footnotesize GRB distributions of $\alpha$ for the analyzed subsamples in 3 considered luminosity ranges: 
$\log L_a < 47.25$ (red), $47.25 < \log L_a < 48.2$ (green) and $\log L_a > 48.2$ (blue). In each $\Delta \alpha$ bin 
the data for the different luminosity ranges are plotted as (shifted within the considered bin) separate color bars. }
   \label{fig:4}
\end{figure}

In the attempt to evaluate the trend in Fig. \ref{fig:2} we decided to compare distributions of $\alpha$
plotted for three luminosity ranges with equal numbers of GRBs: a low luminosity range -- $\log L_a < 47.25$, 
a medium range -- $47.25 < \log L_a < 48.2$ , and a high range -- $\log L_a > 48.2$.
Normalized cumulative distributions function $CDF$ ( $CDF(x) \equiv \sum_0^{x} (1/ N)$, where summing 
includes all GRBs with $\alpha < x$  and $N$ is the number of GRBs in the considered sample) approximates the cumulative 
probability function in the $\alpha$ space.
Below, in Fig. \ref{fig:3} we present these functions in the 3 analyzed luminosity ranges 
for the whole GRBs sample, as well as for long GRBs, short GRBs and XRFs subsamples.
Comparison of the red and blue CDF distributions presented in Fig. \ref{fig:3} convincingly 
(maybe less convincingly for the short subsample) supports the existence of the ($\log L_a$, $\alpha$) 
correlation. The same systematics for all analyzed GRB samples 
with lower luminosity events show tendency to slower light curve decay. The most luminous GRBs 
(blue lines) seem to grow faster to unity with their smaller $\alpha $ scatter. This result is 
also confirmed by the Kolmogorov-Smirnov (KS) test. The test applied for low (red line) and high (blue line) luminosity 
GRB distributions along the $\alpha$ coordinate (Fig. \ref{fig:4}) shows 
that it is highly unlikely, with $P=0.01$, that both distributions are drawn randomly from the same population.
As regards the XRFs and short GRBs subsamples, the available number of elements is too low for establishing
reliable statistical results, however the distributions of brighter GRBs seem to show the same tendency 
to be centered at higher $\alpha$ values than the dimmer ones.

\vspace{0.1cm}

\vspace{0.1cm}

There is no well understood universal recipe yet for differentiating the physical properties of the GRB source 
from observational data, but the existence of $\log L_a$ versus $\log T_a^*$ correlation reflects the presence of 
$\sim$ uniformly varying properties of GRB progenitor in the plateau phase. If these 
properties would be the GRB progenitor mass
and/or its angular momentum, then different external medium profiles could be expected around the exploding massive star, 
where the afterglow related shock propagates. Therefore, it may be expected to detect more clear 
dependence between the afterglow luminosity and the $\alpha$ index for GRBs analyzed within a limited range of $T_a^*$. 
To proceed, in analogy to the above analysis for selected luminosity ranges, we study relative distributions of GRB subsamples
along the LT distribution in three ranges of the decay index:
the ``slowly" decaying light curves with $0.53<\alpha <1.23$, the ``intermediate" ones with $1.23< \alpha <1.54$ and the ``fast" 
decaying light curves with $1.54<\alpha <3.41$. With such selection each subsample has the same size.

\begin{figure}[htbp]
\centering
\includegraphics[width=1.0\hsize,angle=0,clip]{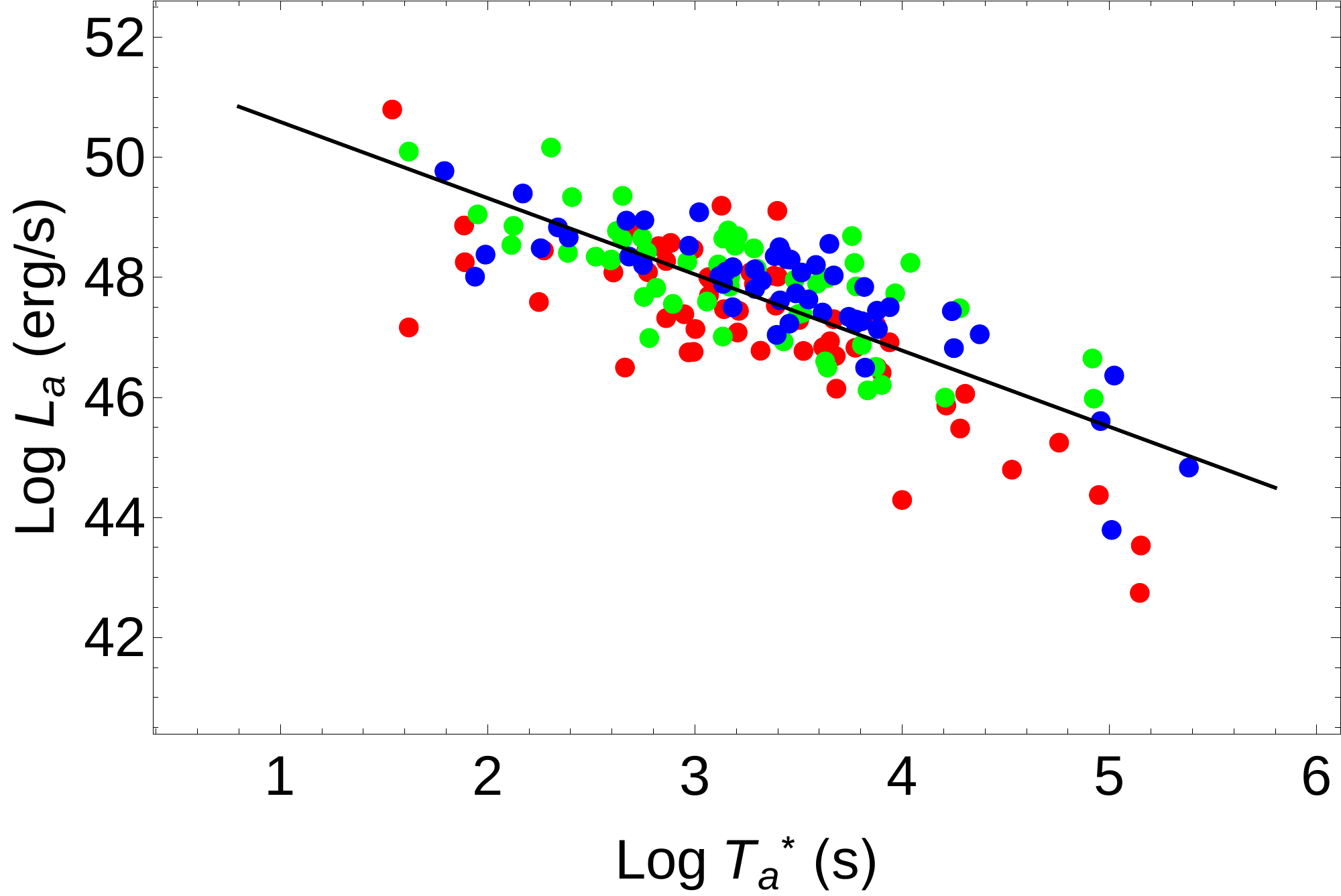}\hspace{1.1cm}
   \caption{\footnotesize Distribution of the long GRBs+XRFs subsample on the ($\log L_a$, $\log T_a^*$) plane 
   for the three selected $\alpha$ subsamples: $0.53<\alpha <1.23$ (red), $1.23< \alpha <1.54$ (green) and  
  $1.54<\alpha<3.41$ (blue). The black line represents the LT correlation line fitted for all presented GRBs.}
   \label{fig:5}
\end{figure}

\begin{figure}[htbp]
\centering
\includegraphics[width=1\hsize,angle=0,clip]{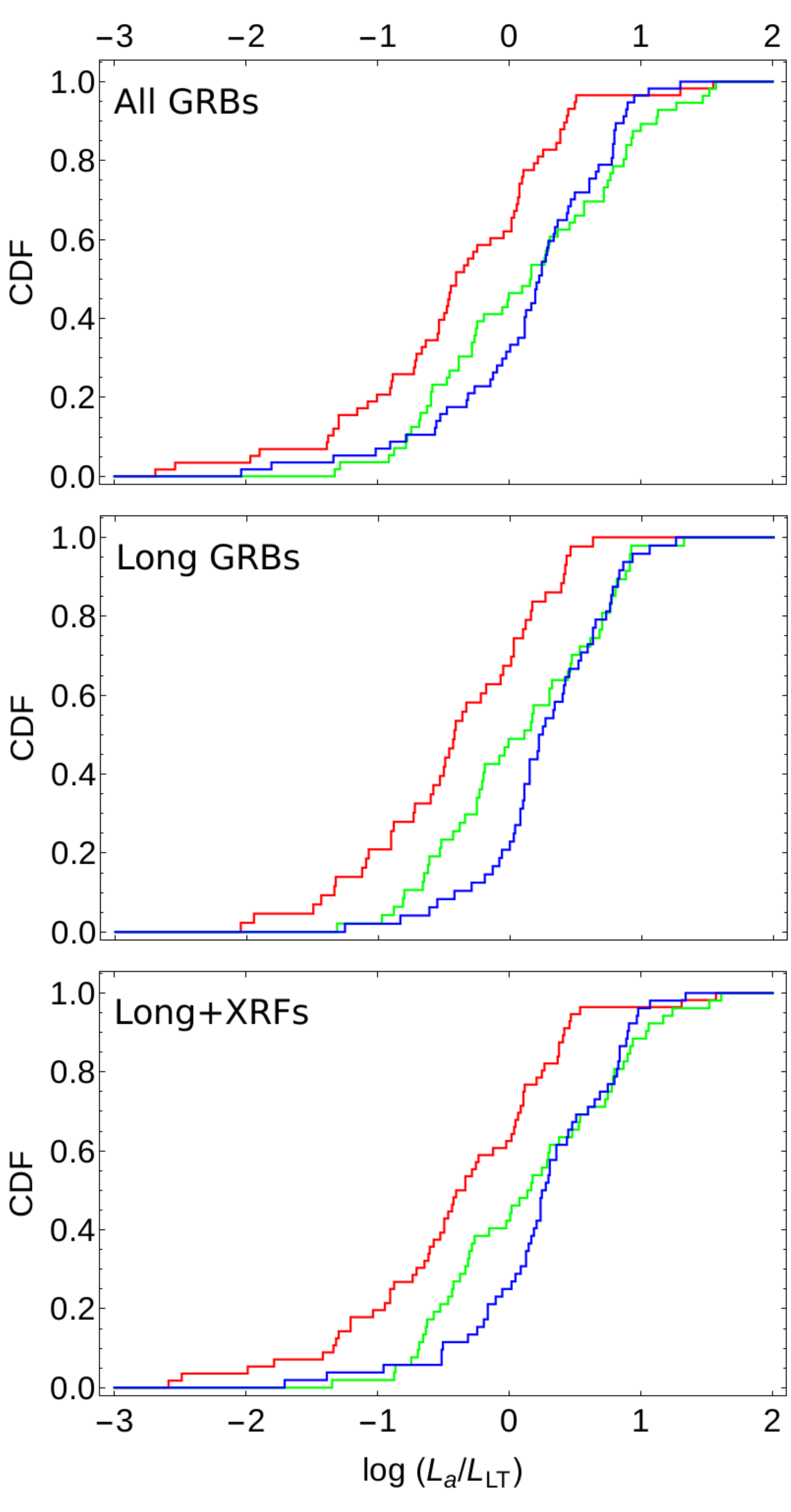}
   \caption{\footnotesize Normalized cumulative distributions function $CDF$ versus  $\log (L_a/L_{LT})$ for the analyzed 
   GRB subsamples in 3 considered $\alpha$ ranges: $0.53<\alpha <1.23$ (red), $1.23< \alpha <1.54$ (green) and 
   $1.54<\alpha<3.41$ (blue).}
   \label{fig:6}
\end{figure}

\begin{figure}[htbp]
\centering
\includegraphics[width=1\hsize,angle=0,clip]{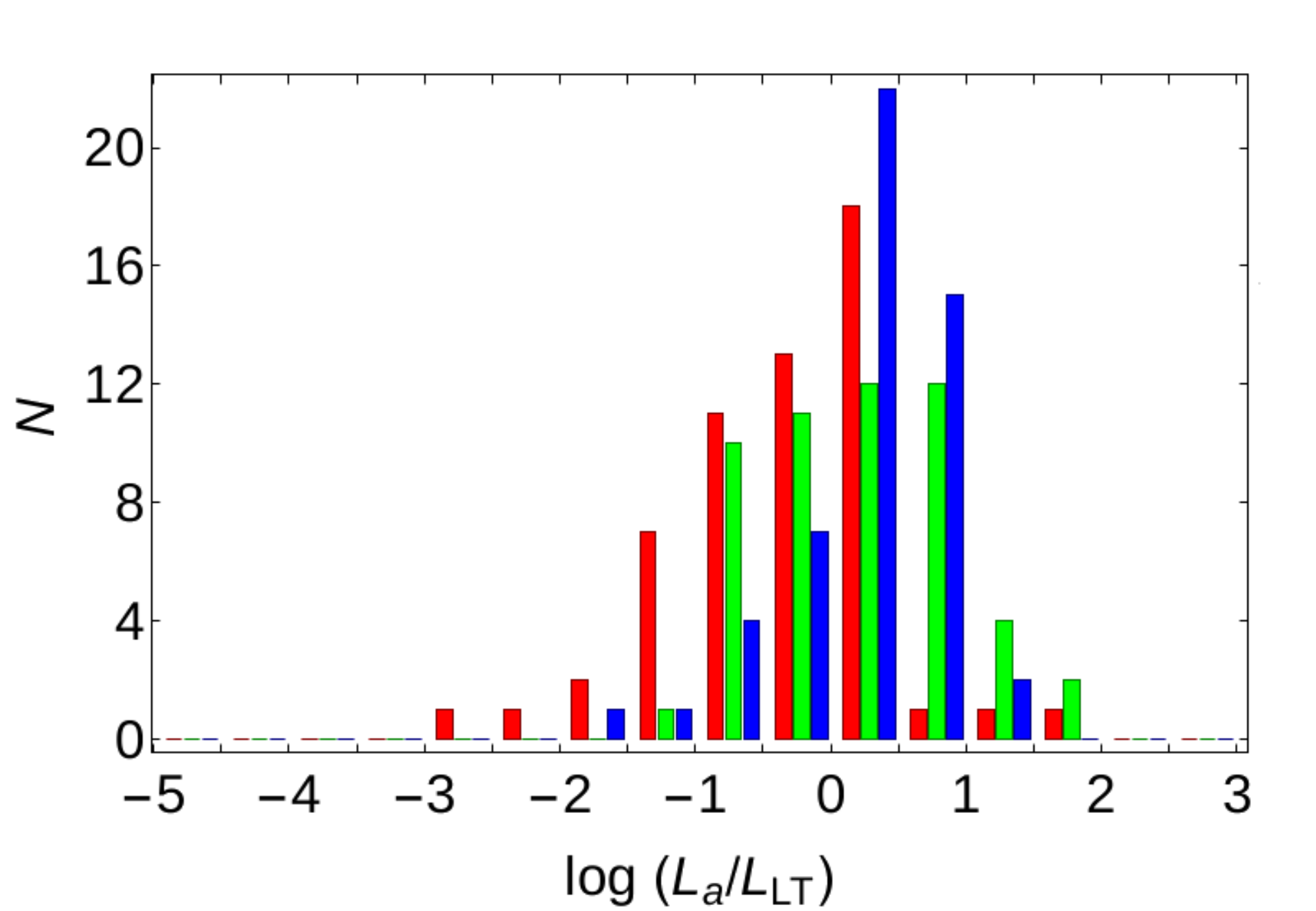}
   \caption{\footnotesize Distributions of $\log (L_a/L_{LT})$ for the long GRBs+XRFs subsamples, for the three $\alpha$ ranges:
   $0.53<\alpha <1.23$ (red), $1.23< \alpha <1.54$ (green) and $1.54<\alpha<3.41$ (blue).}
   \label{fig:7}
\end{figure}

Inspection of Fig. \ref{fig:5}, presenting the considered long GRBs+XRFs $\alpha$-subsamples distributed along the
LT correlation line, clearly reveals separation among these distributions. This behavior is also well visible in the 
normalized cumulative distribution functions plotted in Fig. \ref{fig:6}. 
In particular, in Fig. \ref{fig:6} the considered samples (all, long GRBs and long GRBs+XRFs) are presented with respect to the 
ratio of the GRB afterglow luminosity $L_a(T_a^*)$ to the respective luminosity $L_{LT}(T_a^*)$ at the fitted correlation line: 
in the logarithmic scale $\log (L_a/L_{LT})$ = $\log L_a$ - $\log L_{LT}$~. 
A significant trend between the relative luminosity, $\log (L_a/L_{LT})$, and $\alpha$ is visible in Fig. \ref{fig:7}, 
leading, e.g. 
to negligible KS probability, $P = 1.4\times10^{-6}$, that the low and high $\alpha$ subsamples are 
randomly drawn from a single GRB population.

\begin{figure}[htbp]
\centering
\includegraphics[width=1\hsize,angle=0,clip]{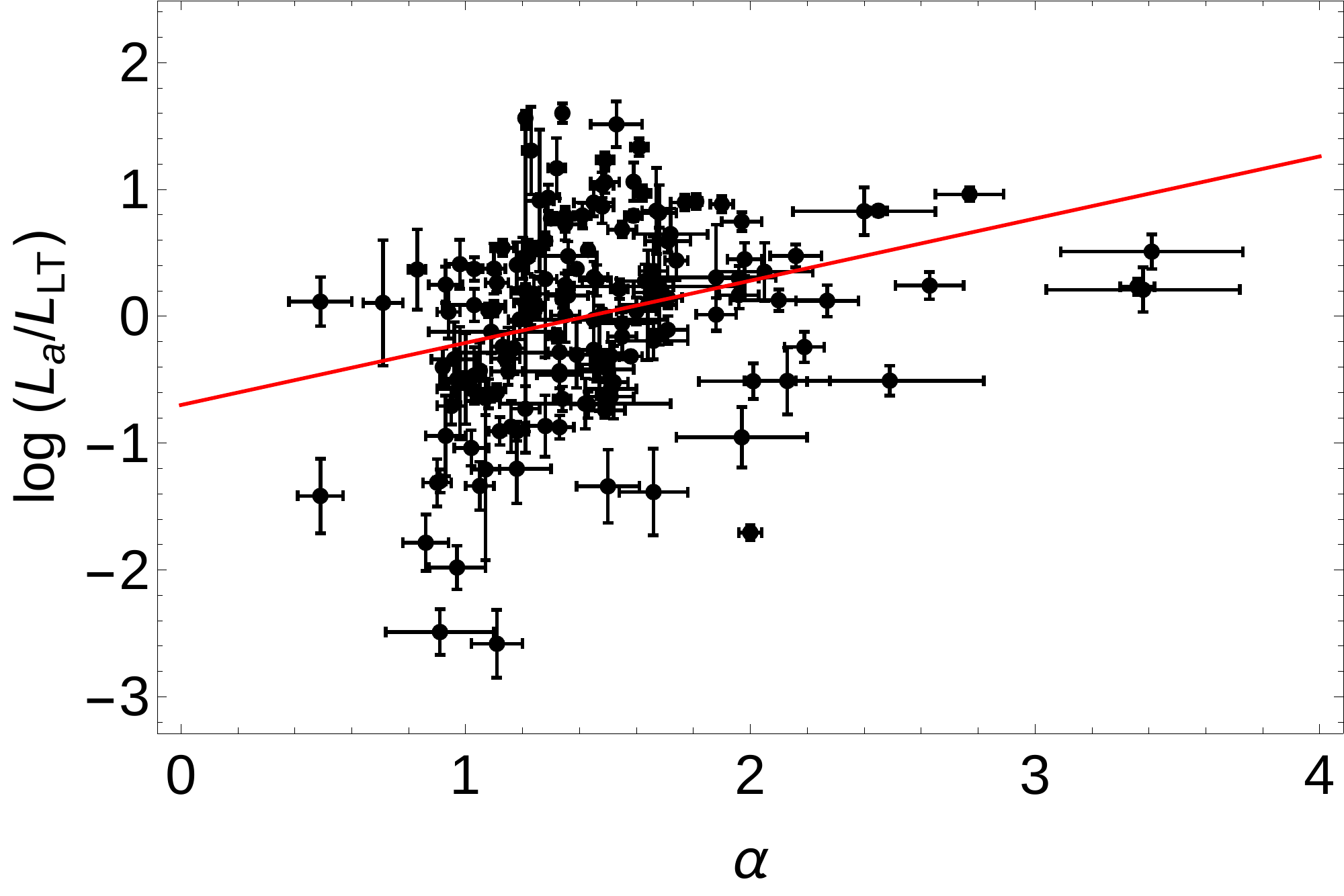}
   \caption{\footnotesize Distribution of the long GRBs+XRFs subsample on the ($\alpha$, $\log (L_a/L_{LT})$) plane.
   The red line presents the fitted correlation (See Eq. 1).}
   \label{fig:8}
\end{figure}

In Fig. \ref{fig:8} we present the distribution of $\log (L_a/L_{LT})$ versus the $\alpha$ index for the long GRBs+XRFs 
subsample. A linear fit for this distribution is 

\begin{equation}
 \log L_a - \log L_{LT} = (0.49\pm 0.17)\,  \alpha - (0.70 \pm 0.25) \quad ,
\end{equation}

\noindent
with Spearman correlation coefficient $\rho = 0.36$, and the probability for random occurrence $P = 10^{-10}$.
Using the long GRB subsample only, the correlation has a slightly smaller slope (0.42) with $\rho=0.36$
and $P=7.7\times10^{-10}$. This correlation shows
the observed tendency -- with respect to the LT correlation line -- 
for higher afterglow luminosity to have steeper light curve decay.

\begin{figure}[htbp]
\centering
\includegraphics[width=1.0\hsize,angle=0,clip]{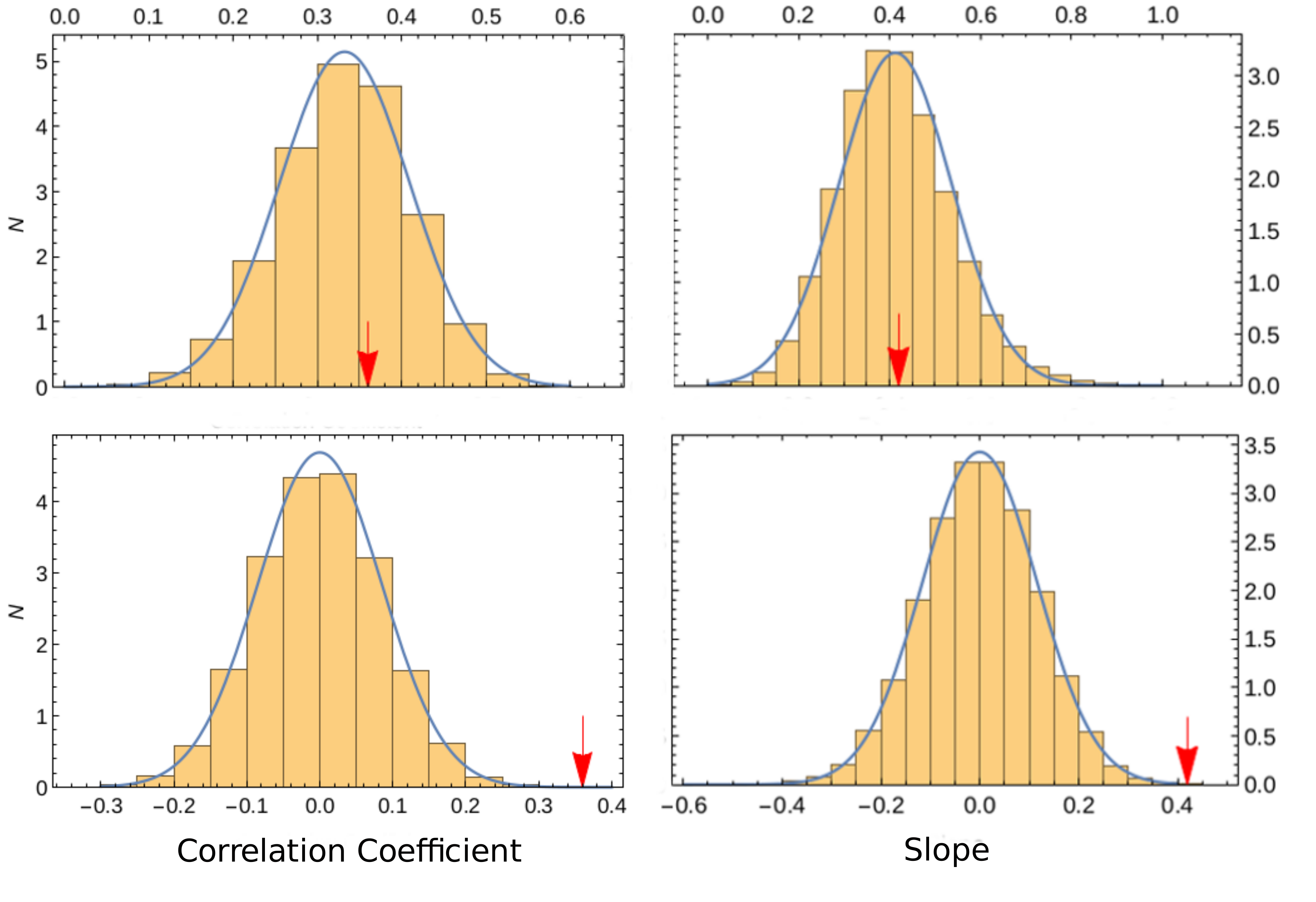}
\caption{\footnotesize The results of the Monte Carlo simulations for the long subsample: 
the normalized distributions of the correlation coefficient $\rho$ and the slope $a$ of the 
$\log (L_a/L_{LT})$ vs. $\alpha$ correlation for bootstrapped 
simulations including random data scatter within the measurement
error ranges (upper panels) and the analogous distributions with
the $\alpha$ parameters separately randomly drawn from the sample
(lower panels). With red arrows we indicated values obtained in the above analysis of the original data.}
   \label{fig:*}
\end{figure}

To better evaluate the errors of the  parameters fitted for the analyzed GRB sample, we decided to perform an 
additional statistical 
analysis using the Monte Carlo modeling of the data with $3\cdot 10^4$ simulations in each case. 
For each GRB we consider parameters $L_a$, $T_a^*$, and $\alpha$ to have 
Gaussian distributions around the fitted values, with the distribution width given by the respective 1 $\sigma$ uncertainty.
Then, we randomly selected from the considered GRB sample - using a bootstrap procedure - 
the samples to be analyzed, where each GRB parameter was drawn from the respective Gaussian 
distribution. For each randomly created data sample we derived the correlation coefficient and the correlation slope 
by fitting the respective correlation $\log (L_a/L_{LT})$ vs. $\alpha$. 
As presented in the upper panels of Fig. \ref{fig:*}, the simulations fully confirm reality of the derived 
correlation. We find that within the measurement errors existing 
correlation coefficient should be approximately between $0.2 < \rho < 0.5$ (mean value 0.35) and the fitted 
$\log (L_a/L_{LT})$ vs. $\alpha$ correlation should have an inclination $0.3 < a < 0.5$ (mean value 0.41), 
in agreement with the fitted errors in Eq. (1). 

Using similar simulations as above, it can also be independently checked the possibility of randomly obtaining the studied
$\log (L_a/L_{LT})$ vs. $\alpha$ correlation if no systematic relation of $\alpha$ in respect to $L_a$ and $T_a^*$ exists. 
We performed such analysis by randomly drawing 
samples using the procedure above, again within the bootstrap scheme, 
but with separately drawing pairs of parameters $L_a$ and $T_a^*$ from the GRB sample,
and the $\alpha$ values from the sample of these values for the considered GRBs. Such
procedure removes any correlation between $\alpha$ and other GRB parameters 
in the sample and shows that the possibility of randomly obtaining the correlation 
coefficient found in the real data is negligible (Fig. \ref{fig:*}, lower panels). 

\section{Final discussion and Conclusions}

Analysis of the fitted afterglow power-law temporal decay indices for the subsample of long GRBs+XRFs reveals a weak 
trend towards a steeper decay phase for higher afterglow luminosity $L_a$. The trend turns into a significant correlation 
if we consider GRB afterglow 
luminosity scaled to the one expected from the fitted LT correlation, for a given GRB afterglow plateau end time $T_a^*$. 
As different $T_a^*$ values result from varying properties of the GRB sources, the present analysis can be used to get new 
insight into physical nature of such sources. 

\subsection{Theoretical models}

It is worth to note the attempts in the literature to provide physical interpretation of $\log L_a$ versus $\alpha$ 
relation. We can refer to such model presented by \cite{hascoet14} (see also \cite{genet07}) in order to relate the 
considered $\alpha$ parameter to the microphysics of the reverse shock emission.
In the model of Hascoet et al. the energy deposition rate, ${\dot E}_T$, in the GRB afterglow, varied in time `$t$' with a
power-law dependence on the Lorentz gamma factor, $\Gamma (t)$:

\begin{equation}
\label{eq_lfDistPeak}
{\dot E}_{\rm T}(\Gamma(t))= \left\lbrace\begin{array}{cl}
{\dot E}_*\left(\frac{\Gamma(t)}{\Gamma_*}\right)^{-q}\ \ \ {\rm for}\ \ \Gamma(t)>\Gamma_*\\
{\dot E}_*\left(\frac{\Gamma(t)}{\Gamma_*}\right)^{q^{\,\prime}}\ \ \ {\rm for}\ \ \Gamma(t)<\Gamma_*\\
\end{array}\right.
\end{equation}

\noindent
where the energy scale ${\dot E}_*$ is determined by the total energy injected in the afterglow phase; $q$ and $q^{\, \prime}$ are
the power-law indices for the time dependence of the energy injection rate. 
In this model the $q$ parameter constraints the shape of the plateau 
phase, while $q^{\, \prime}$ carries information about the light curve temporal decay index after the plateau. 
The characteristic value of $\Gamma_*$ sets the duration of the plateau. 
With the assumed power-law radial distribution of the medium surrounding the GRB progenitor, the Lorentz factor evolves as

\begin{equation}
 \Gamma(t)=\Gamma_*\left( \frac{t}{T^*_a}\right)^{-\gamma},
\end{equation}

\noindent 
where, e.g., $\gamma = 3/8$ for a uniform medium and $\gamma = 1/4$ for a stellar wind. 
Hascoet et al. derived the light curve temporal decay indices before ($\alpha_1$) and after ($\alpha_2$) the break at 
the end of the plateau phase as

\begin{equation}
\label{eq_rel_indices}
\left\lbrace\begin{array}{ll}
\alpha_1=\gamma q-1\\
\alpha_2=-\gamma q^{\,\prime}-1   
\end{array}\right.
\end{equation}

\noindent
so that the flat plateau phase should be present for $q \approx 1/\gamma$, i.e. close to $q=8/3$ in the uniform medium and
$q=4$ in the wind case.
In the presented example \cite{hascoet14} considered the temporal decay index after the plateau $\alpha_2 = -1.5$, 
leading to $q^{\, \prime}= 1/(2\gamma)$ and the parameters of the central engine energy deposition in the late afterglow stage 
$q^{\, \prime} = 4/3$ and 2, for
the uniform medium and the wind case, respectively. It should be noted that this example uses the $\alpha_2$ value very 
close to the mean value of our distribution, $\alpha_{mean}=1.4 \pm0.3$, as  visible in the upper panel of Fig. \ref{fig:1}. 

We should remark here that the observed large scatter in the $\alpha$ distribution seems to be difficult to be explained
only by variations of the source radial density profile, influencing the shock propagation. Therefore, we consider the 
present discussion only as an example of the study based on the $\alpha$ parameter measurements. 

\subsection{GRB standardization} 

As the second aspect of the present study, we consider a possible usage of the measured afterglow light curve 
$\alpha$ for physically differentiating the observed GRBs. 
E.g., the revealed $\log(L_a/L_{LT})$ versus $\alpha$ correlation can be used to search for the procedure which
could enable the standardization of GRBs and eventually to reveal a new 
cosmological standard candle. 
As an {\em illustrative toy model}, for such approach we introduce a procedure for GRBs resembling the one used for the
standardization of SN Ia light curves by using the so called Phillips relation between the peak magnitude and the ``stretching 
parameter'' \citep{phillips}.
In an attempt to scale GRBs with different temporal decay indices $\alpha$ to the standard source properties, 
we define the standard GRB as the one characterized by the value of the temporal decay index 
$\alpha_0 = 1.4$, approximately the mean value of the distribution presented in Fig. \ref{fig:1}. 
Further, we postulate that the expected ``standardized" GRB luminosity, $L_{a,0}$, can be derived using its measured 
decay index $\alpha$ and by scaling it to $\alpha_0$ using Eq. 1:

\begin{equation}
\log L_{a,0} = \log L_a + 0.49 \, (\alpha_0 - \alpha) \quad .
\end{equation}

\noindent
This procedure applied for all the events in the analyzed subsample of long GRBs+XRFs results in only a slight
increase in the LT correlation coefficient absolute value, from
$\rho = -0.72$ for the original ($\log L_a$, $\log T_a^*$) data to $\rho_0 = -0.76$ for the modified distribution ($\log L_{a,0}$, 
$\log T_a^*$). When using the long GRBs subsample only the increase in the correlation coefficient is even smaller.
This increase in the correlation coefficient is obtained by the fits for quite 
different shapes of the afterglow light curves and thus different quality of available GRB parameters $L_a$, $T_a^*$ and $\alpha$. 
More detailed analysis of this standardization procedure, with careful consideration of the afterglow light
curves for the selected events, is in progress now.

\vspace{0.45cm}
\noindent \footnotesize{Acknowledgements.} This work made use of data supplied by the UK Swift Science Data Centre at the 
University of Leicester.
The work of RDV and MO was supported by the Polish National Science Centre through  the grant DEC-2012/04/A/ST9/00083. M.G.D is 
grateful to the Marie Curie Program, because the research leading to these results has received funding from the European 
Union Seventh FrameWork Program (FP7-2007/2013) under grant agreement N 626267.

\end{document}